# Heat transfer in sound propagation and attenuation through gas-liquid polyhedral foams

Yuri M. Shtemler[*], Isaac R. Shreiber

Department of Mechanical Engineering

Ben-Gurion University of the Negev

P.O. Box 653, Beer-Sheva 84105, Israel

**Abstract**

A cell method is developed, which takes into account the bubble geometry of polyhedral foams, and provides for the generalized Rayleigh-Plesset equation that contains the non-local in time term corresponding to heat relaxation. The Rayleigh-Plesset equation together with the equations of mass and momentum balances for an effective single phase inviscid fluid yield a model for foam acoustics. The present calculations reconcile observed sound velocity and attenuation with those predicted using the assumption that thermal dissipation is the dominant damping mechanism in a range of foam expansions and sound excitation frequencies.

## 1. Introduction

Gas-liquid foams have attracted attention for years because of their unusual physical properties and wide range of applications [1-6]. In particular, foam acoustics offers a variety of possible applications: noise-level reduction of blasts when foam fills the blast chamber, and of turbine engine exhaust by means of foam injection, foam injection; monitoring the foam drive processes for increasing the recovery of oil etc. In a way similar as in usual gases velocity and attenuation of sound waves in foams may correlate with the corresponding characteristics of combustion waves, and thus determine the regimes and mechanisms of the foam combustion. Of the large body of investigations carried out on gas-liquid foams, only a small fraction is devoted to sound, shock and combustion waves in foams [7-15]. Short reviews of foam acoustics are given in [11, 13]. The foam response to sound and weak shock waves demonstrates low dispersion of sound velocity along with high sound attenuation [7, 11]. In closely related systems such as bubbly liquids theoretical investigation of acoustic waves is reduced to their study in the effective single-phase media with the help of cell methods [16-18]. The cell methods allow to relate the pressure to the density of the effective single-phase fluid (or, equivalently, to the bubble radius) by the generalized Rayleigh-Plesset equation for bubble dynamics. Together with the equations of mass and momentum balances for an effective single phase fluid Rayleigh-Plesset equation yields the model for bubbly liquid acoustics. The similar approaches have been applied to highly wet foams accounting for several approximations of the viscous resistance force between the gas and liquid phases [9-11]. However, for sufficiently low sound excitation frequencies thermal dissipation may be identified as the dominant damping mechanism over the viscous friction [13, 14]. The conventional cell methods [9-11], and the earlier distributed analog method [4] are based on the explicit hydrodynamic solution around a spherical bubble within the infinite layer of the incompressible liquid. They assume the radius of spherical cells to be much larger than the bubble size [9-11, 14]. This is rather invalid for moderately wet foams which are well approximated by the packings of pentagonal dodecahedral bubbles with thin liquid films as faces and long plateau borders (PBs) as edges (Fig. 1).

_________________
[*]Corresponding author, e-mail address: shtemler@ bgu.ac.il; FAX: +972-8-6472813;
Tel: +972-8-6477055

The current work adopts the main developments in the spherical-cell methods to the polyhedral foams taking into account the bubble geometry by the proper choice of cells.

## 2. Cell method for polyhedral gas-liquid foams

Below the following main qualifications are adopted: the length of the sound waves is fairly larger than the foam bubble size and smaller than the foam sample size; thermal dissipation is the dominant damping mechanism over the viscous friction; the disjoining pressure is nedligible small; gravity effects and resulting bubble size inhomogeneity over the height of the foam sample are neglected; coalescence of foam bubbles and gas diffusion through the liquid films are ignored; the foam gas is compressible and heat-conductive; while the foaming liquid is incompressible and isothermal. The volume of the foam films is fairly less than the PBs one, and hydroconductivity of the thin liquid films is negligibly small compared with that of the PBs.

Let us consider the flow within a PB capillary which arises as a response to sound wave effect. A one-dimensional hydraulic approach to be taken for the mass and momentum (neglecting inertial and viscous terms in Darcy-like approximation) balances yields for the flow of an inviscid incompressible and isothermal liquid in PB:

$$\frac{\partial A}{\partial t} + \frac{\partial (Au_l)}{\partial z} = 0, \qquad \frac{\partial P_l}{\partial z} = 0, \qquad (0 < z < L/2). \tag{1}$$

The dynamic and kinematic conditions at the bubble-liquid interface are as follows:

$$P_l = P_g - 2\sigma/R_c, \qquad u_l = \partial R_c/\partial t \qquad \text{at} \qquad z = 0. \tag{2}$$

Here $u_l(z,t)$ and $P_l(z,t)$ are the velocity and pressure averaged across the PB; $t$ is time; $z$ is the local coordinate along the axis of the PB with the length and cross-section area, $L(t)$ and $A(z,t)$; $P_g$ is the gas pressure within the foam bubbles; $\sigma$ is the surface tension; and $R_c$ is the local curvature radius of PBs, in particular, the local radius of foam bubbles at the node. Due to the symmetry of the system the foam stress tensor or, equivalently, the foam pressure, $P_f$, is defined as the liquid pressure at the boundary between two neighbouring cells where the liquid velocity is equal to zero (similar to the spherical cell methods for bubbly liquids):

$$P_f \equiv P_l, \qquad u_l = 0 \qquad \text{at} \qquad z = L/2. \tag{3}$$

In neglect of the spatial variations of both gas pressure and temperature within the bubbles, the gas state equation and the bubble energy equation read [16-18]:

$$\frac{P_g V_g}{T_g} = \frac{P_{g0} V_{g0}}{T_{g0}}, \qquad \frac{dP_g}{dt} + \frac{3\gamma P_g}{R_b}\frac{dR_b}{dt} = -\frac{3(\gamma-1)}{R_b} q_b, \tag{4}$$

where the gas phase consists of equal-size bubbles with the volume equivalent spherical radius $R_b$; $T_g$ is the gas temperature; subscript 0 denotes the initial equilibrium parameters of the foams; $q_b$ is the heat transfer rate at the bubble interface; $\gamma$ is the adiabatic constant.

A gas within foam bubbles obeys the ideal gas law, $P_g = (\gamma-1)c_{Vg}T_g\rho_g$; $c_{Vg}$ is the specific heat capacity of gas; and $V_g(t) = \alpha_g/\rho_f$ is the specific volume of gas

$$V_g = 4/3\,\pi R_b^3 N_b. \tag{5}$$

Here and below the following definitions for two-phase mixtures are used [17]:

$$\alpha_m = V_m \rho_f, \quad X_m = V_m \rho_m, \quad V_f \rho_f = 1, \quad \rho_f = \alpha_g \rho_g + \alpha_l \rho_l, \quad \alpha_g + \alpha_l = 1, \tag{6}$$

$\rho_g, \rho_l, \rho_f$ are the gas, liquid and foam densities; $\alpha_g$, $\alpha_l$ are the gas, liquid contents; $m = g, l$.

Neglecting the bubble coalescence and the liquid losses from the foam sample yields for the bubble number per unit mass, $N_b$, and mass concentration of the liquid, $X_l$,

$$N_b = const, \qquad X_l = const. \tag{7}$$

Newton's law of cooling is assumed to be valid for the heat transfer rate per unit of contact area between a bubble and the surrounding liquid (Appendix A):

$$q_b = (T_g - T_l)h_b. \tag{8}$$

Here variations of the liquid temperature during the bubble oscillation are negligible because of high heat capacity of the liquid; $T_l = T_{g0} \equiv const$; $h_b$ is the surface heat transfer coefficient.

To close the system the bubble geometry is taken into account by the well-known relations between the PB structural (geometrical) parameters, $L$, $A$, $R_c$, and foam expansion, $K_f$:

$$L(t) = \sqrt[3]{\frac{4\pi}{3\beta_1}} R_b(t), \quad A(t) = \beta_2 R_c^2(t), \quad A(t) = \frac{2\beta_1}{n} \frac{L^2(t)}{K_f(t) - 1}, \quad K_f(t) = \frac{1}{\alpha_l(t)}. \quad (9)$$

Here the PB is a capillary bounded by three neighboring bubbles forming triangular cross-sections and ends with two bubbles (Fig. 1); the length of PB is suggested to be much larger its cross-section size, $n = 20$ is the vertex number of the dodecahedron, $\beta_1 = (15 + 7\sqrt{5})/4 \approx 7.6$, $\beta_2 = (\sqrt{3} - \pi/2) \approx 0.162$. The first of Eqs. (9) relates the PB length (length of the dodecahedron edge), $L$, to the bubble radius, $R_b$. The second of Eqs. (9) relates the PB cross-section area $A$ to the curvature radius $R_c$ of a PB (restricted by three mutually contacting cylinders of the radius $R_c$). The third of Eqs. (9) follows from the liquid content definition if the volume occupied by the total liquid films is negligibly small compared to the total PBs volume:

$$\alpha_l = \frac{V_l}{V_f} \approx \frac{ALn/2}{\beta_1 L^3 + ALn/2}. \quad (10)$$

Bubble radius $R_b(t)$ and foam expansion $K_f(t)$ may be expressed through foam density, $\rho_f$:

$$R_b \approx \sqrt[3]{\frac{3}{4\pi N_b \rho_f}} \equiv R_{b0} (\frac{\rho_{f0}}{\rho_f})^{1/3}, \qquad K_f = \frac{\rho_l}{X_l \rho_f} \equiv K_{f0} \frac{\rho_{f0}}{\rho_f}. \quad (11)$$

Let us suggest the approximation of moderately wet foams is valid, i.e.

$$\frac{\rho_{f0}}{\rho_l} = \frac{\rho_{g0}}{\rho_l} + \frac{1}{K_{f0}} + O(\frac{1}{K_{f0}^2}), \qquad \frac{\rho_{g0}}{\rho_l} \lesssim K_{f0}^{-1} \equiv \alpha_{l0} << 1. \quad (12)$$

In this limit both terms should be preserved in the right hand of Eq. (12). Equations (9)-(12) relate the current values of foam expansion and structural parameters to the foam density:

$$\frac{K_f}{K_{f0}} = (\frac{\rho_f}{\rho_{f0}})^{-1}, \quad \frac{R_b}{R_{b0}} = \frac{L}{L_0} \approx (\frac{\rho_f}{\rho_{f0}})^{-1/3}, \quad \frac{R_c}{R_{c0}} \approx (\frac{\rho_f}{\rho_{f0}})^{1/6}, \quad \frac{A}{A_0} = (\frac{\rho_f}{\rho_{f0}})^{1/3}. \quad (13)$$

The PB curvature radius is much less than the bubble radius that is of the order of the PB length:

$$\frac{R_{c0}}{R_{b0}} = \frac{B_1}{\sqrt{K_{f0}}}, \qquad \frac{L_0}{R_{b0}} = B_2, \qquad \frac{A_0}{R_{b0}^2} = \frac{B_3}{K_{f0}}, \quad (14)$$

$$B_0 = \sqrt[6]{\frac{16\pi^2 \beta_1}{9}}, \quad B_1 = \sqrt{\frac{2}{\beta_2 n}} B_0 \approx 1.8, \quad B_2 = \frac{B_0}{\sqrt{\beta_1}} \approx 0.8, \quad B_3 = \beta_2 B_1^2 \approx 0.52. \qquad (15)$$

Then, heat transfer coefficient may be estimated as follows (Appendix A, [19, 20]):

$$h_b = 1/3 B\, k / R_{b0}, \quad B = 3\pi/(16\beta_2) N_e B_2 Nu, \qquad (16)$$

$N_e = 30$ is the number of PBs which bound a foam bubble; $k$ is the gas thermal conductivity; $Nu \approx 3.66$ is the Nusselt number ($\beta_2 \approx 0.162, B = 3\pi/(16\beta_2) N_e B_2 Nu \sim 400$). This expression for $h_b$ is valid for quasi-steady state systems, for which the heat relaxation time, $\tau$, and the sound excitation frequency, $\omega$, satisfy:

$$\tau\, \omega < 1. \qquad (17)$$

To relate parameters of the flow within a PB with the foam structural parameters let us assume that the structural parameters are independent of the longitudinal coordinate. This is valid for high lengths of the PB compared with the characteristic size of the PB cross section. Then in the hydraulic approximation adopted here, the former of Eqs. (1), after integration with respect to $z$, yields that the liquid velocity contains arbitrary function of time $M_1(t)$

$$u_l(z,t) = -z A^{-1} dA/dt + M_1(t), \qquad 0 < z < L/2. \qquad (18)$$

Hence, the liquid velocity cannot satisfy the kinematic boundary conditions at the bubble interface in Eqs. (2) and simultaneously at the cell boundary in Eqs. (3). It may be shown in the limit of moderately high foam expansions that the kinematic condition in Eqs. (2) may be omitted everywhere outside a negligibly thin boundary layer in the node vicinity. Then the dynamic condition in Eqs. (2), (3) together with Eqs. (1) give the liquid velocity and pressure:

$$u_l(z,t) = (1 - 2\frac{z}{L}) L \frac{\dot{R}_c}{R_c}, \qquad P_l(z,t) = -\frac{2\sigma}{R_c} + P_g, \qquad (19)$$

$$\frac{P_g}{P_{g0}} = (\frac{\rho_f}{\rho_{f0}})^\gamma \frac{1}{E}\{1 + \frac{1}{\tau}\int_0^t (\frac{\rho_f}{\rho_{f0}})^{1/3-\gamma} E\, dt'\}, \qquad E = \exp(\frac{1}{\tau}\int_0^t (\frac{\rho_f}{\rho_{f0}})^{-2/3} dt) \qquad (20)$$

Here the non-local term in Eq. (20) arises from the explicit solution of the bubble energy equation (4) after substitution of Newton's law (8); $\tau$ is the heat relaxation time.

Equations (3), (19) yield the generalized Rayleigh-Plesset equation for bubble dynamics:

$$P_f = P_g - \frac{2}{R_{b0}} \sigma_f (\frac{\rho_f}{\rho_{f0}}), \qquad \frac{\sigma_f}{\sigma} = \frac{\sqrt{K_{f0}}}{B_1} (\frac{\rho_f}{\rho_{f0}})^{-1/6}, \qquad (22)$$

where $\sigma_f$ denotes the effective coefficient of the foam interfacial tension.

## 3. Governing equations for foam acoustics

Let us represent any quantity, $f$, near the equilibrium state as follows:

$$f = f_0 + \delta f + .... +, \qquad f_0 \equiv const, \qquad \delta f \ll 1, \qquad (23)$$

where $f = \{\rho_f, P_f, v_f, K_f, ...\}$; subscript 0 denotes the equilibrium state, $v_{f0} \equiv 0$; $\delta f$ denotes small perturbations. Thus in the linear approximation, gas-liquid foams can be described by the general inviscid hydrodynamic equations of mass and momentum balances together with the generalized Rayleigh-Plesset equation (similar to [21]):

$$\frac{\partial \delta \rho_f}{\partial t} + \rho_{f0}\frac{\partial \delta v_f}{\partial x} = 0, \quad \frac{\partial \delta v_f}{\partial t} = -\frac{1}{\rho_{f0}}\frac{\partial \delta P_f}{\partial x}, \quad \delta P_f = \delta P_g + \frac{\sigma_{f0}}{3R_{b0}\rho_{f0}}\delta\rho_f, \quad (24)$$

$$\delta P_g = \gamma \frac{P_{g0}}{\rho_{f0}}\delta\rho_f - \tau^{-1}\frac{P_{g0}}{\rho_{f0}}\int_0^t \exp(-\frac{\theta}{\tau})\delta\rho_f(t-\theta)d\theta, \quad \frac{\sigma_{f0}}{\sigma} = \frac{1}{B_1}K_{f0}^{1/2}. \quad (25)$$

Introducing the effective values of high and low-frequency sound velocities of foams $C_f^{(\infty)}, C_f^{(0)}$, ($C_f^{(\infty)} > C_f^{(0)}$), and omitting for brievity subscript 0, the relation for $\delta P_f$ may be rewritten as:

$$\delta P_f = C_f^{(\infty)2}\delta\rho_f - \tau^{-1}(C_f^{(\infty)2} - C_f^{(0)2})\int_0^t \exp(-\theta/\tau)\delta\rho_f(t-\theta)d\theta. \quad (26)$$

Equation (26) is similar to that in [21] which has the infinite upper limit in the integral term,

$$C_f^{(\infty)} = \sqrt{\frac{\gamma P_{g0}}{\rho_{f0}} + \frac{\sigma_{f0}}{3R_{b0}\rho_{f0}}} \equiv C^{(0)}\sqrt{\gamma + \Pi_C}, \quad C_f^{(0)} = \sqrt{\frac{P_{g0}}{\rho_{f0}} + \frac{\sigma_{f0}}{3R_{b0}\rho_{f0}}} \equiv C^{(0)}\sqrt{1 + \Pi_C},$$

$$C^{(0)} = \sqrt{\frac{P_{g0}}{\rho_{f0}}} \equiv \frac{C_g^{(0)}}{\sqrt{1 + K_{f0}^{-1}\rho_l/\rho_{g0}}}, \quad C_g^{(0)} = \sqrt{\frac{P_{g0}}{\rho_{g0}}}, \quad \Pi_C = \frac{\sigma_{f0}}{3R_{b0}P_{g0}},$$

where $\Pi_C$ is the capillary criterion.

Applying the Laplace and Fourier transforms with parameters $s$ and $\lambda$ with respect to $t$ and $x$ to any function $f(x,t)$ results in

$$\hat{f}(\lambda, s) = \int_{-\infty}^{+\infty}\int_0^\infty f(x,t)\exp(-st)\exp(-i\lambda x)dtdx. \quad (27)$$

Setting the initial conditions to zero and introducing excitation frequency, $\omega$, instead of the parameter of the Laplace transform, $s = -i\omega$, yield

$$\delta \hat{P}_f(\omega) - C_f^{(0)2} \delta \hat{\rho}_f(\omega) = -\frac{i\omega\tau}{1-i\omega\tau}(C_f^{(\infty)2} - C_f^{(0)2})\delta \hat{\rho}_f(\omega), \tag{28}$$

where parameters of pressure deviation in the right-hand side of Eq. (28) may be related to the second viscosity of foams. Eq. (28) yields the Mandelshtam-Leontovich relation by for a complex sound velocity [22]:

$$C_f^2(\omega) = \frac{\delta \hat{P}_f(\omega)}{\delta \hat{\rho}_f(\omega)} = C_f^{(0)2} - \frac{i\omega\tau}{1-i\omega\tau}(C_f^{(\infty)2} - C_f^{(0)2}). \tag{29}$$

The complex wave number $\lambda = \omega / C_f$ corresponds to an exponential planar wave

$$f(x,t) = \hat{f}(\lambda,\omega)\exp(i\lambda x - i\omega t) \equiv \hat{f}(\lambda,\omega)\exp(-\lambda_i x + i\lambda_r x - i\omega t). \tag{30}$$

The wave propagates along the positive direction of axis $x$ with real excitation frequency $\omega$ and wave number $\lambda_r > 0$ ($\lambda = \omega / C_f$, $\lambda = \lambda_r + i\lambda_i$). Then, Eq. (29) yields the well-known dispersion relation for sound relaxation in single-phase gas media [22]:

$$\lambda = \frac{\omega}{C_f^{(0)}}\sqrt{\frac{1-i\omega\tau}{1-i\omega\tau\, C_f^{(\infty)2}/C_f^{(0)2}}}. \tag{31}$$

Due to surface tension smallness at the presence of surfactants, $\Pi_C = 0$ everywhere below.

## 4. Comparison of modeling results with experimental data

The acoustic characteristics of foams vs. the heat relaxation time, $\tau$, are presented in Fig. 2, the maximal value of the sound attenuation is achieved at $\tau_{max} \approx (\gamma\omega)^{-1}$, while the sound velocity varies within the range $C_f^{(0)} < C_f < C_f^{(\infty)}$. Since $\tau$, in Eq. (21), depends on the initial radius of the foam bubble, $R_{b0}$, for which the relaible experimental data may be unavailable, first $\tau$ is considered as the fitting parameter and then $R_{b0}$ is estimated. Fitting the modeling to measured sound attenuation, $\lambda_i$, yields two values $\tau_1$ and $\tau_2$ (Fig. 2a). The smaller of two values $\tau$ is selected in order to better satisfy the quasi-steady state condition (17), $\omega\tau < 1$: $\omega\tau = \{0.23, 0.08, 0.71\}$ at $\omega/(2\pi) = \{300, 600, 300\}$ $Hz$ for experimental data adopted from [7, 11, 12], respectively (see Fig. 3a, data from [12] have been obtained by using the empirical correlation for $\lambda_i = 0.036 K_f^{-0.625}(\omega/2\pi)^{1.25}$). Then the effective values of $R_{b0} \approx \{1.1, 0.45, 1.9\}\, 10^{-3} m$ are calculated by Eq. (21) (with $\kappa \approx 2.2 \cdot 10^{-5}$ $m^2/s$ for air) and compared with the values $R_{b0} \approx \{1.0, 0.23, 1.0\}\, 10^{-3} m$ reported in [7, 11, 12], respectively, which start a series of experiments within the range of foam expansions. Thus, our predictions for $R_{b0}$ are in fair agreement with [7], but overestimate the values in [11, 12]. This is explained noting that in [7] the bubble size is held constant during all the experiments by rotation of the foam sample, thus preventing the escape of liquid from the foam, while the foam expansion is varies independently of $R_{b0}$. Whereas in [11, 12] both the bubble size and foam expansion rise with time due to the gravity effects, and then the starting values $R_{b0}$ reported in [11, 12] are less than the effective values for $R_{b0}$ in the model prediction. The predicted sound velocities are in a qualitative agreement with experimental values [7, 12], in particular, they demonstrate a weak dispersion, but overestimate them (see Fig. 3b, where error bars represent the standart errors assosiated with measurement precision). Experimental data [7] demonstrate the relatively high errors typical for the wave velocity measurements by pressure sensors in relaxing media.

## 5. Summary and conclusions

The cell method is developed for polyhedral foams which provides for the generalized Rayleigh-Plesset equation that relates the foam pressure to density and contains the non-local in time term corresponding to the heat relaxation. Using of Newton's law of cooling for the heat transfer between the foam bubbles and surrounding liquid films allows to derive an explicit form of the "equation of state" for gas-liquid foams. The general inviscid hydrodynamic equations of mass and momentum balances together with the generalized Rayleigh-Plesset equation yield the model for acoustics in polyhedral gas-liquid foams. Application of the Laplace and Fourier transforms to the equations for foam acoustics leads to the well-known dispersion relation for thermal relaxation of sound in effective single-phase gas media. Neglecting additionally input of surface tension the modeling results are compared with experimental data. The available experimental data are in fair agreement with the model prediction for sound velocity and attenuation. This demonstrates that acoustic characteristics of polyhedral foams are mainly governed by heat-transfer effects in a range of foam expansions and sound excitation frequencies. In general case for quantative comparison of the modeling and experimental data measurements of bubble size are needed simultaneously with measurements of the foam expansion. On the other side based on the resulting mathematical model the effective bubble size in the polyhedral foam can be controlled using measurements of sound attenuation.

**Acknowledgements**

The financial support of the Israel Science Foundation under grant 278/03 is gratefully acknowledged.

**Nomenclature**

$A$, $A_{pb}$  cross-section area and lateral area of plateau border

$A_b$  surface area of foam bubble

$B_k$  coefficients of the foam structural parameters, $k=0,1,2,3$

$C_f^{(\infty)}, C_f^{(0)}$  high and low-frequency sound velocities of foams

$c_{pg}$  specific heat capacity of gas

$h_b$, $h_{pb}$  effective coefficients of the heat transfer for bubble surface and plateau border

$P_m$  foam pressure of gas, liquid, foam, , $m=g,l,f$

$q_b$, $q_{pb}$  heat transfer rates at the bubble and plateau border surfaces

$K_f$  foam expansions

$k$  gas thermal conductivity

$L$  length of plateau-border

$N_b$  bubble number per unit mass

$N_e$  is the number of plateau borders which bound the foam bubble

$Nu$  is the Nusselt number

$n$  dodecahedron' vertex number

$R_b$  radius of equivalent spherical bubbles

$R_c$  curvature radius of plateau borders

$s$  parameter of the Laplace transform

$t$  time

$T_m$  temperature of gas, liquid, $m=g,l$

$u_l$  liquid velocity in plateau border

$V_m$  specific volume of gas, liquid, foam, $m=g,l,f$

$v_f$  foam velocity

$X_m$  mass concentration of gas/liquid, $m=g,l$

$x$  coordinate along the foam sample

$z$  local coordinate in plateau border

*Greek letters*

$\alpha_m$  gas, liquid volume content, $m=g,l$

$\beta_k$  coefficients of the foam structural parameters, $k=1,2,3$

$\delta f$  small perturbations of $f$

$\gamma$  adiabatic constant

$\lambda=\lambda_r+i\lambda_i$  complex wave number

$\Pi_C$,  acoustic-capillary criterion

$\rho_m$  density of gas, liquid, foam, $m=g,l,f$

$\sigma, \sigma_f$  surface tension and foam effective surface tension

$\tau$  heat relaxation time

$\omega$  excitation frequency

**Appendix A. Estimation for the heat transfer coefficient at the bubble surface**

Equating the total heat flux through the surface of the PBs contacting with a foam bubble to the total heat flux homogeneously distributed over the foam bubble surface yields:

$$A_b \mid q_b \mid = N_e A_{pb} \mid q_{pb} \mid, \tag{A.1}$$

$A_b \approx 4.2\pi R_{b0}^2$ is the area of a foam bubble, $A_{pb} = \pi L R_{c0} \equiv \pi B_2 R_{b0} R_{c0}$ is the lateral area of the PB contacting with the bubble (Eqs. (15)), $N_e = 30$ is the number of the dodecahedral edges which bound the foam bubble. The heat flux of the liquid filled PB may be estimated as [19]:

$$q_{pb} = (T_l - T_g) h_{pb}, \quad h_{pb} \approx k Nu / R_{eq}, \tag{A.2}$$

where $R_{eq} = 2 A_0 /(\pi R_{c0}/2) \approx (4\beta_2/\pi) R_{c0}$ is the equivalent hydraulic radius of the PB; $k$ is the gas thermal conductivity, and the Nusselt number may be estimated by $Nu \approx 3.66$ for fully developed laminar flow in the equivalent cylindrical capillary with constant surface temperature. Rigorously speaking expression (A.2) for $h_{pb}$ is valid for the steady-state systems [19, 20], so relation (17) should be satisfied. Substituting (A.2) into (A.1) yields Eqs. (8), (16):

$$q_b = (T_g - T_l) h_b, \quad h_b = 1/3 B k / R_{b0}, \quad B = 3\pi/(16\beta_2) N_e B_2 Nu \approx 319.2. \tag{A.3}$$

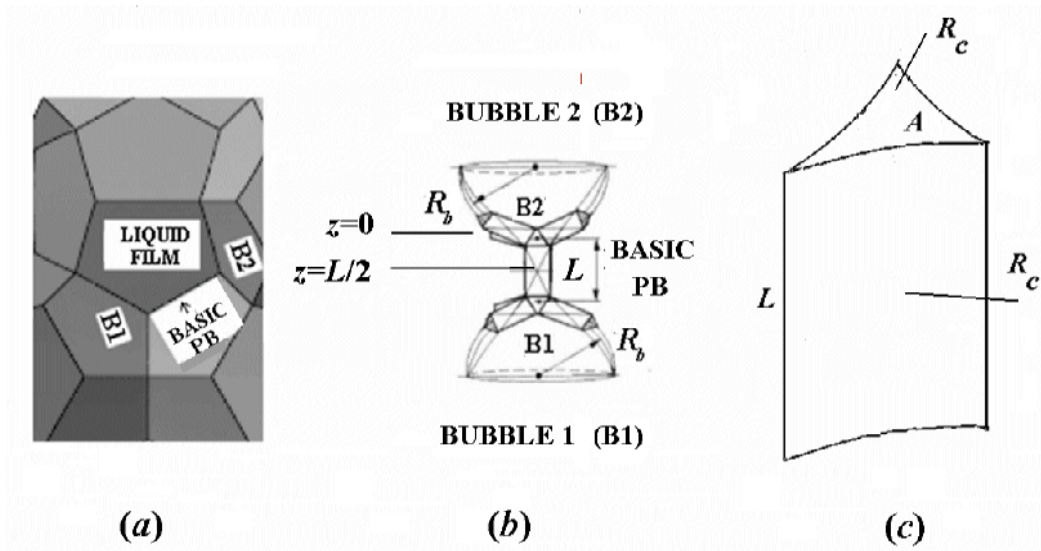

Fig 1. Cell model for gas-liquid polyhedral foams (schematic)

(a) foam sample; (b) basic cell unit bounded by bubble surface and PB half; (c) basic PB.

B1, B2 are the bubbles ending the basic PB; $R_b$ is the radius of the equivalent spherical bubble;

$L$ $R_c$, $A$ are the PB's length, curvature radius and cross-section area.

Y. M. Shtemler, I. R. Shreiber.
Heat transfer effects in sound propagation and attenuation through gas-liquid polyhedral foams

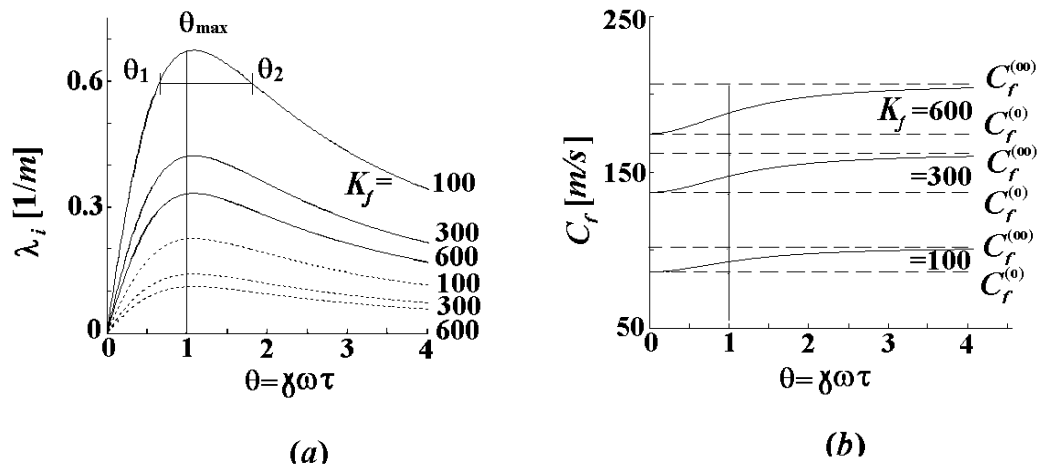

Fig. 2.  Sound attenuation (*a*) and velocity (*b*) vs. the scaled relaxation time $\gamma\omega\tau$

(modeling results for air-water foam at several foam expansion $K_f$).

Fig. 2a: $\omega/(2\pi) = 750 Hz$ (solid lines); $\omega/(2\pi) = 250 Hz$ (point lines).

Fig. 2b: sound velocity, $C_f$ (solid lines);

low and high frequency sound velocities, $C_f^{(\infty)}$ and $C_f^{(0)}$ (dashed lines).

Y. M. Shtemler, I. R. Shreiber.
Heat transfer effects in sound propagation and attenuation through gas-liquid polyhedral foams

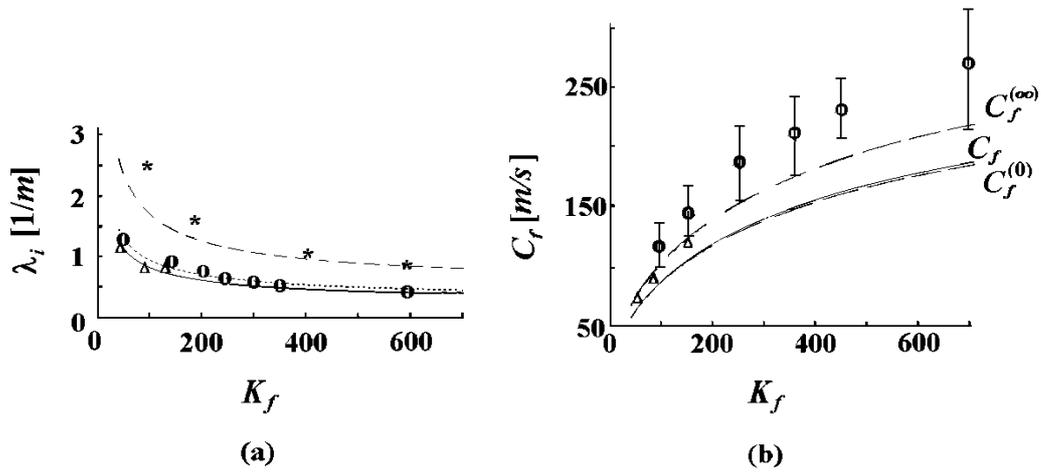

Fig. 3. Sound attenuation (*a*) and velocity (*b*) vs. foam expansion for air-water foams

(comparison of experimental data and present modeling), $\pi_c = 0$.

$\Delta$, $\bigcirc$, $*$ -- experimental points adopted from [11,7,12], respectivelly.

Fig. 3a: $\omega/(2\pi)=645Hz$; $\omega\tau = 0.08$ (solid line);

$\omega/(2\pi)=300Hz$; $\omega\tau = 0.23$ and $0.71$ (pointed and dashed lines).

Fig. 3b: sound velocity $C_f$ (dispersionless for $50Hz \leq \omega/(2\pi) \leq 750Hz$, solid line);

high and low-frequency sound velocities $C_f^{(\infty)}$ and $C_f^{(0)}$ (dashed lines).

Y. M. Shtemler, I. R. Shreiber.
Heat transfer effects in sound propagation and attenuation through gas-liquid polyhedral foams